\documentclass{article}

\usepackage{microtype}
\usepackage{graphicx}
\usepackage{subcaption}
\usepackage{booktabs} %

\usepackage{hyperref}

\usepackage[accepted]{icml2026}

\usepackage{amsmath}
\usepackage{amssymb}
\usepackage{mathtools}
\usepackage{amsthm}
\usepackage{cuted}
\usepackage{caption}
\usepackage[capitalize,noabbrev]{cleveref}

\theoremstyle{plain}

\theoremstyle{definition}

\theoremstyle{remark}

\usepackage[textsize=tiny]{todonotes}

\icmltitlerunning{Position: EU AI Act's Research Exemptions Can Break the Publication Norms of Major AI Conferences}

\begin{document}

\twocolumn[
  \icmltitle{Position: EU AI Act's Research Exemptions Can Break the Publication Norms of Major AI Conferences}

  \icmlsetsymbol{equal}{*}

  \begin{icmlauthorlist}
    \icmlauthor{Alina Wernick}{yyy}
    \icmlauthor{Kristof Meding}{yyy}

  \end{icmlauthorlist}

  \icmlaffiliation{yyy}{CZS Institute for AI and Law, University Tübingen, Germany}

  \icmlcorrespondingauthor{Alina Wernick}{alina.wenick@uni-tuebingen.de}

  \icmlkeywords{Machine Learning, AI Act, Research }

  \vskip 0.3in
]

\printAffiliationsAndNotice{}  %

\begin{abstract}
The EU has become one of the vanguards in regulating the digital age. A particularly important regulation in the Artificial Intelligence (AI) domain is the 2024 enacted EU AI Act. The AI Act specifies --- due to a risk-based approach --- various obligations for providers of AI systems. These obligations, for example, include a cascade of documentation and compliance measures, which represent a potential obstacle to science. But do these obligations also apply to AI researchers? This position paper argues that, indeed, the AI Act's obligations could apply in many more cases than the AI community is aware of. Moreover, we argue that the AI Act is drafted in a manner that may unwillingly disrupt the scientific publication practices of the AI research community, with a focus on model and system release. We contribute the following: 1.) We offer a high-level roadmap for AI researchers to evaluate whether they need to comply with the AI Act 2.) We explain with everyday research examples why the AI Act applies to AI research. 3) We analyse the exceptions of the AI Act's applicability AI research and offer visual tool for researchers to navigate the AI Act's complex system or research exceptions 4.) We establish a position the AI Act's research exceptions fail to account for current AI research conventions, as publishing AI research may void the research exceptions of the Act. 5.) We propose changes to the AI Act to provide more legal certainty for AI researchers and give two recommendations for AI researchers to reduce the risk of not complying with the AI Act. We see our paper as a starting point for a discussion between policymakers, legal scholars, and AI researchers to avoid unintended side effects of the AI Act.
\end{abstract}

\section{Introduction}
Discriminatory and harmful uses of AI have caught regulators' attention \citep{IAPP, whitecase}, with the EU passing a pioneering Act on Artificial Intelligence \citep{AIA}, which on the one hand, seeks to protect EU citizens from the risks of AI, and on the other hand fosters the adoption and development of trustworthy AI on the EU market. Unfortunately, despite its commitment not to interfere with scientific research (Rec. 25), the AI Act has not been drafted to account for the AI research practice of publishing alongside scientific papers on platforms like GitHub, Hugging Face or Colab. As a consequence, AI researchers who publish their research artifacts at conferences such as ICML, NeurIPS, or ICLR may be held liable under the AI Act. 

Why should the international AI and ML community care about the applicability of the AI Act to their research? To begin with, the current ICML Research Ethics\footnote{See \url{https://icml.cc/Conferences/2026/ResearchEthics}.} requires AI Act compliance. As we explain below, applying the AI Act to research may result in significant compliance obligations for researchers, which they lack the time, budget, and expertise to meet. Researchers would be required to adhere to the same regulatory requirements that apply to companies like Meta, Google, or Anthropic. These requirements can create obstacles for researchers. The obstacles exist regardless of whether researchers are working within the EU or outside of it, such as in the USA. Similar to the General Data Protection Regulation (GDPR), the AI Act has an extraterritorial scope of application. Therefore, we see that even outside the European Union, the AI Act might also affect researchers in two ways. First, in simplified terms, if an AI system operates in or targets the European market ((Art. 2), the AI Act applies regardless of the origin of the provider or deployer. Second, we envision a so-called Brussels effect\cite{bradford2020brussels} of the AI Act beyond the European Union, in which ML researchers should also engage with their respective policymakers. Due to the scope ofapplication of the AI Act and the resulting obligations, it is, in our view, crucial for researchers to consider its applicability to their research practices. Where a research practice is not protected by the AI Act's research exceptions, researchers must follow all compliance regulations of the AI Act. Failing to follow the AI Act's obligations is not a trivial matter, as fines for non-compliance with rules on AI are up to 35,000,00€ (Art. 99 (3)).

We deploy the legal methodology of interpreting the law to discern how the AI Act applies to AI research, to systematize its complex rules on research exceptions and to substantiate their meaning \cite{legalmethods}. To the extent it is possible, we seek to clarify its ambiguities with legal interpretation \cite{chynoweth2008legal}. For the legally interested reader, we summarised the previous literature on the AI Act in section \ref{ref:previousLiterature} in the appendix. %

\textbf{Our position is that A) the AI Act's obligations apply to AI research in many more cases than the AI community and the legislator were aware of B).} \textbf{This leaves researchers in uncharted waters and facing severe legal challenges and liabilities, that risk disrupting established research practices.} Our contribution to these challenges:
\begin{itemize}
  \setlength\itemsep{0.0em}
    \item We will give a high-level introduction of the AI Act in \ref{IntroAIAct}, targeting Machine Learning researchers with no prior knowledge of the AI Act, and show the severe implications of AI Act applicability. Furthermore, we will provide examples of AI research that fall into the categories of AI systems regulated by the AI Act.
    \item We will analyze in-depth the exceptions of the AI Act in favour of scientific research, product-oriented research, and open source, with a focus on model and system release. Furthermore, we explain why, in many cases, the AI Act does not account for AI research practices in Section \ref{secException}.
    \item Finally, we will reflect on legislative measures and interpretations that could mitigate the challenges the AI Act poses on ML research, see Section \ref{sec:Certanty}.
\end{itemize}
\vspace{-0.3cm}
\section{Research and AI Act}

\label{IntroAIAct}
We will first provide a high-level introduction to make the implications of the AI Act understandable to all researchers. 
\vspace{-0.3cm}
\subsection{A Primer on the AI Act}
The AI Act\footnote{All legal citations refer to the AI Act, if not otherwise stated} of the European Union came into force on August 1, 2024 2024. The AI Act creates a uniform legal framework for the development, placing on the market, putting into service, and the use of AI systems in the EU (Rec. 1 ). The AI Act seeks to promote the uptake of human-centric and trustworthy AI, ensuring a high level of protection of health, safety, and fundamental rights, and against the harmful effects of AI while supporting innovation, including science and research \& development (R\&D) (Art. 1, Rec. 25). To fulfil its regulatory objectives, the Act sets numerous compliance obligations on the providers of the AI systems that should cover the lifetime of the system (Recs. 69, 71). This raises the question of whether AI researchers also need to comply with the AI Act.

To this end, the researcher must first assess whether they are dealing with an AI system or an AI model regulated by the AI Act, and whether, through their actions, they qualify as providers of these. We will cover both aspects next.
\vspace{-0.2cm}
\subsection{Categories and Processing of AI systems}
\label{sec:Categories}
The AI Act imposes compliance obligations across multiple categories of AI systems, with the most extensive requirements applying to high-risk AI systems and general-purpose AI models. Unfortunately, differentiating between what is an AI system and what is an AI model is challenging and is subject to legal uncertainty. This is because the AI Act defines the AI system very broadly (Art. (3)(1)) and in a manner that does not correspond to the language adopted in the AI research community \citep{Nolte2024}. We will discuss this challenge in more detail in the Appendix \ref{sec:ModelSystem} and get back to the issue in section 4. What is essential is that combining AI models with other components, such as a user interface, qualifies them as AI systems (Recital 97; \cite{Nolte2024}) regulated by the AI Act. Similarly, offering demos of their research output (for example, on a website) can be seen as an AI system.

\textbf{Main categories in the AI Act: prohibited AI systems, high-risk AI systems, and GPAI models} 

To be on the safe side, AI researchers must evaluate whether their research concerns AI categories regulated by the AI Act: prohibited AI systems, high-risk AI systems (HRAI), or GPAI models.\footnote{Please also note that there is a fourth special category, certain AI systems in Article 50, with specific transparency obligations, which are not covered in this work.}

Firstly, research may involve practices that are prohibited by the AI Act, such as certain subliminal techniques, systems for exploiting vulnerabilities of a natural person or a specific group, social scoring systems, as well as systems for emotion or biometric recognition, as well as certain facial databases (Art. 5). Examples for this type of research could be \textit{Labeled Faces in the Wild }\citep{huang2008labeled}\footnote{The most popular face-recognition database at Papers with Code \url{https://paperswithcode.com/datasets?task=face-recognition}} or \citet{kosinski2021facial} who published a paper on facial recognition technology. 

Secondly, the AI Act lays out detailed, technically and administratively laborious compliance obligations for HRAI systems. HRAI systems include those that represent products or safety components to products covered by the New Legislative Framework, such as machinery and medical devices (Art. 6(1); Annex III), and systems that are deployed for (Art. 6 (2); Annex III) biometrics; critical infrastructure; educational and vocational training; employment; worker's management or access to self-employment, access to and enjoyment of essential private services and essential public services and benefits; law enforcement; migration, asylum and border control management and administration of justice and democratic processes. Examples from science can also be found for high-risk systems \citep{yan2024practical,yigit2024critical,tambe2019artificial}. Thus, if the AI Act applies to science, current research might be classified as high-risk systems, triggering a cascade of obligations.
In contrast to the example of an AI system in the high-risk area mentioned above --- which might sound a bit distant to reseachers --- the AI Act also regulates GPAI models. As a reminder, a GPAI model is an AI model that is trained with a large amount of data using self-supervision at scale, displays significant generality, and is capable of competently performing a wide range of distinct tasks (Art. 3(63)). 
Examples of GPAI might include models such as GPT \citep{brown2020language} or Gemini \citep{team2023gemini}. Additionally, more open models like LLaMA \citep{dubey2024llama}, DeepSeeks-R1 \citep{guo2025deepseek}, Falcon \citep{almazrouei2023falcon}, or Bloom \citep{le2023bloom} might also be considered GPAI models. \\

\textbf{Activities that trigger compliance obligations}

If AI researchers are dealing with one of the regulated systems just explained, they must review whether they are likely to engage in actions that trigger AI Acts' compliance obligations for providers over the course of their research activity. The term provider refers to an actor that develops an AI system or a model. The key actions that trigger the provider's obligations are ``placing on the market" and ``putting into service" (Art. 3(3)).

Generally, an AI system or model covered by the AI Act must comply with it the moment it is made available on the Union Market for the first time (i.e. placed on the market) (Art. 3 (9)) \citep{Commission_2022}. The term \textit{making available on the market} means the supply of an AI system or a GPAI model for distribution or use on the Union market in the course of a commercial activity, whether in return for payment or free of charge (Art. 3 (10)). In principle, offering the possibility to download or use an AI system over the cloud could trigger the applicability of the AI Act, see \citet{Commission_2022}. As discussed in more detail in Section 3.9, the meaning of the concept of 'commercial activity' is not unambiguous \cite{finck2026eu}. 

\textit{Putting into service} means the supply of an AI system for the first use directly to the deployer or for own use in the EU for its intended purpose (Art. 3(11)). Putting into service also covers AI systems developed for in-house use \citep{Commission_2022}, within an AI research institution. It can thus be established that many usual activities of AI researchers could, in principle, qualify as putting into service or making available on the market. 
 
\subsection{Legal Burdens of the AI Act to Researchers}
Where AI researchers suspect that they are working on prohibited or HRAI systems or GPAI models regulated by the AI Act, they must assess whether their research activity triggers compliance obligations or liabilities under the AI Act. Providers must ensure that the HRAI systems and GPAI models they place on the market or put into service comply with all the requirements of the AI Act (Arts. 3 (3) 16, 53, 55). 
We will discuss examples of these severe requirements for high-risk system and GPAI models in the following.

For high-risk systems, a risk management system must be implemented (Art. 9). This risk management system must be planned in advance and maintained throughout the entire life cycle. For research, this would also include maintenance after publication. According to Art. 10, data governance practices need to be implemented. This includes documentation of design choices, personal data, and bias mitigation techniques. In addition, accuracy, robustness, and cybersecurity issues must be addressed (Art. 15). This covers technical redundancy, backup, and fail-safe plans. All requirements must be documented throughout the system's entire lifespan (Art. 11 and Art. 12). These obligations will be detailed in harmonized standards, see \citet{kilian2025european} for more details.
 
For GPAI models, similar requirements are established (Art. 53 ). However, these requirements are currently being defined by the Code of Practice. The current draft of the Code of Practice includes roughly 50 pages of requirements, for example, on transparency, copyright, and safety. While details are not yet negotiated, many of the rules focus on (internal) compliance \cite{zuiderveen2024non} and checklists. From our perspective, the AI Act’s requirements pose a substantial (documentation) burden for the ordinary ML researcher.

Non-compliance with the rules poses a risk of severe penalties. The AI Act defines in Article 99 and Article 101 fines for non-compliance with the AI Act. These penalties are primarily targeted at companies (e.g., Art. 99 (3)) but can also target public entities such as universities. %
The amount of the fine depends on whether, for example, the ban on prohibited AI systems (Art. 99(3); up to 35,000,000 €) or specific high-risk system regulations (Art. 99 (4) to 15,000,000 €) have been violated. 

\section{Research Exceptions for AI}

It should be noted that AI research concerning unregulated forms of AI does not need to comply with the Act. However, as shown above, AI research may consist of regulated forms of AI: prohibited systems, high-risk systems, and GPAI models, see also Figure \ref{fig:exceptions} for an overview.

\label{secException}
\subsection{A Labyrinth of Exceptions}

At first glance, the AI Act's commitment to the freedom of science and its aim of not undermining R\&D Activity (Rec. 25) seems convincing --- after all, it contains several exceptions in favor of scientific research (Art. 2(6), Rec. 109), r\&d (Arts. 2 (8), 3(63), 57-62) and open-source AI (Arts. 2 (12), 53(3)). In its Digital Strategy, the European Commission considers the AI Act as one of the policies promoting AI research and innovation, highlighting that ``The regulation is not applicable to any activities related to AI research, testing, and development before it is marketed or put into operation"\cite{EUAIDigitalStrategy2025}. Mantelero shares this view, stating ``that the AI Act is largely not applicable to research activities"\cite{Mantelero2025}. 

\begin{figure*} 
    \centering
    \includegraphics[width=\textwidth]{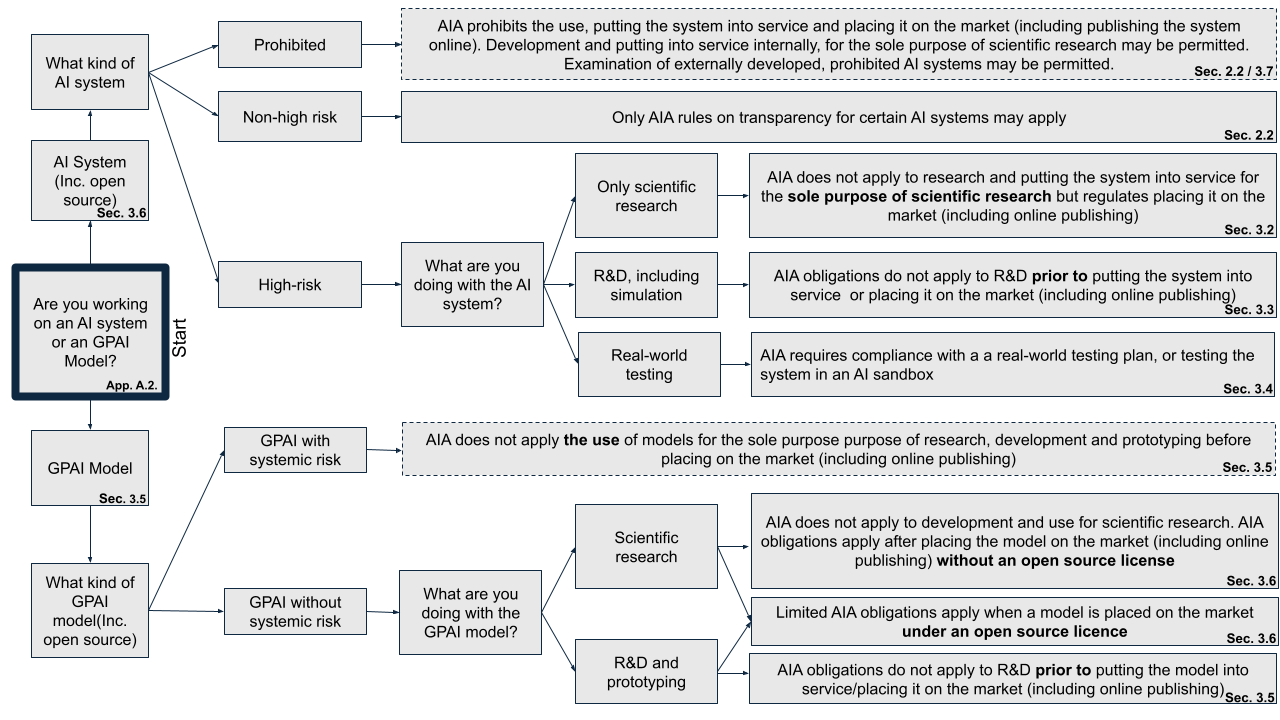}
    \caption{This flowchart illustrates the (simplified) system of AI Act's research-relevant exceptions, which are explained in detail in section \ref{secException}. It analyses whether researchers might need to comply with the AI Act as providers for prohibited and high-risk AI systems and GPAI models. }
    \label{fig:exceptions}
    \vspace{-0.5cm}
\end{figure*}

However, we argue that the research exemptions of the AI Act are not drafted to account for some AI research practices, exposing researchers to a maze of exceptions, and considerable legal uncertainty about their obligations under the AI Act. Scholars have called for clarification of the scope of the AI Act's research exceptions\cite{latil2025regulating}. In response, we devised a flowchart to support AI researchers' sense making of the Act's system research exceptions. (see Figure \ref{fig:exceptions}).\footnote{Please note that the Figure and our study does not account for rules on certain systems subject to transparency obligations (Art. 50) or which concern researchers as deployers of AI systems (Art. 3(4)) We do not reflect on the AI Act's exceptions for AI developed for military, defence, and security purposes (Art. 2(3)).} This system of relevant exceptions for researchers is explained in plain text in the following section, which the Figure \ref{fig:exceptions} cross-references.

Due to the misalignment of these exceptions with AI research practices, the act of publishing an AI system or model may void the research exceptions and trigger compliance obligations --- and legal liability under the Act. Without legal measures and interpretations that create legal certainty for AI research, there is a risk that the AI Act will hinder scientific research and the practice of publishing AI systems or models alongside scientific publications. 

The discrepancy between the expected effect of research exceptions and their fit with the research practices of the AI research community may have originated from the lack of consideration for AI research practices throughout the AI Act's legislative process. The Act's impact on academic research was not reviewed \cite{IAAI2021}, and its earlier drafts were criticized for not containing research exemptions \citep{smuha2021eu, Amsterdampaper, bogucki2022ai}. The Act was refined throughout the legislative process to include the research exceptions analysed below \citep{AIAcomparison}. 

\subsection{Scientific Research} 
\label{sec:SR}
The AI Act establishes an exception for scientific research that covers AI systems or AI models, including their output, \textit{specifically developed and put into service for the sole purpose} of scientific research and development (Art. 2(6); Rec. 25). However, this exception covers only the act of putting the AI system or model into service.

Considering the networked nature of today's research practice with many stakeholders and participants \citep{colonna2023ai}, this exception is very narrow in scope. It covers situations where the AI system is developed for the sole purpose of in-house scientific research or the direct deployment by a research collaborator (Arts. 2 (6); 3(11)). AI systems that may be used to carry out any product-oriented research, development and testing do not fulfil the requirement of ``sole purpose'' (Rec. 25). In practice, the exception is precarious \cite{finck2025search} and it is difficult for researchers to foresee whether and when their research crosses the boundary from scientific research to research and development, especially in research partnerships that may include private companies \citep{colonna2023ai}.

The main question is whether publishing a model belongs to the core research (or scientific) activity (which is not part of placing it on the market or putting it into service). Uploading research artifacts to repositories like GitHub, Hugging Face or Colab may simplify further research and initiate a new development loop (See \citet{gorwa2024moderating} for governance challenges). Please note first that ``making available on the market'' does not require any payment and can be done free of charge (Art. 3(10)). Thus, just because ML research artifacts can be downloaded from repositories without any fee does not imply that they have not been placed on the market (for a discussion of the open-source exception, see below and \cite{Commission_2025}). \\
Uploading code and models could be seen as an integral part of scientific research. Major machine learning conferences such as NeurIPS --- the primary research outlets for machine learning-related research --- encourage researchers to publish their models and code.\footnote{See \url{https://neurips.cc/Conferences/2025/CallForPapers}.} It is the standard (recommended) practice to publish code and models \citep{pineau2021improving} when submitting a paper within the research community.

However, uploading the models and code might not belong to the core research activity. The Oxford dictionary defines research as ``systematic investigation or inquiry aimed at contributing to the knowledge of a theory, topic, etc., by careful consideration, observation, or study of a subject''.\footnote{\url{https://www.oed.com/dictionary/research_n1?tab=meaning_and_use\#25922908}.} Still, one can argue that uploading research artefacts wouldn't be necessary to perform the core ML research activity. According to the wording of the Oxford dictionary, it can be argued that uploading research artefacts is not part of ``systematic investigation or inquiry aimed at contributing to the knowledge''. See also Figure \ref{fig:ML_Workflow} for illustration.

The act of publishing a system, in other words, placing it on the market or making it available on the EU market, falls out of the scope of the scientific research exception.

\subsection{Product-oriented research}
The AI Act has a distinct exception for product-oriented research, testing, and development activity for AI systems and models (Rec. 25). The AI Act does not apply to R\&D prior to the AI system or model being placed on the market or put into service (Art. 2(8)).The exception for product-oriented research seems to protect the internal R\&D activity of commercial AI providers. Similarly, R\&D partnerships are not carved out of the AI Act, instead, the act seeks to offer AI providers with relative freedom to experiment in the R\&D phase, under the condition that the product is compliant when they seek to commercialize it. In practice, complying with the AI Act requires AI providers to start planning for AI Act compliance already in the R\&D phase (See \cite{colonna2023ai}

Here too, product-oriented research at large companies with research units might be considered a borderline case of science \citep{biega2021reviving}. Research units at large companies like Google or Meta perform research similar to that in academic institutions. Although they sometimes have a focus on product-oriented research, it might be less clear how the product-oriented research exception should be interpreted with respect to work conducted at the companies' research-oriented units. Commentators have accepted that the work conducted by units such as Google Health could, under certain circumstances, be regarded to fall under the AI Act's exception for scientific research rather than R\&D \citep{de2021researcher}.

We recognize the necessity of not watering down the obligations of the AI Act for companies by overly generous interpretation of the Act's exceptions in favour of research and product development (Art. 2(6) and 2(8)) \citep{colonna2023ai,kazim2023proposed,bogucki2022ai}. However, the absence of specific rules for publishing AI systems and models in the course of research remains a serious problem. By following the established practices of the field and publishing their research artifacts, AI researchers working in companies and universities risk prematurely triggering the applicability of the AI Act for the entire R\&D project involving the relevant AI system This can occur also where actors in question have genuine intentions to comply with the AI Act's regulations in the event that the AI system in question will be commercialized. This premature imposition of compliance obligations would serve neither research community nor R\&D in the field of AI.

\subsection{Real-Life Testing and Regulatory Sandboxes}
Even when AI researchers benefit from the exceptions for product-oriented research, they must note that the exception does not extend to real-world testing of HRAI (Recs. 25, 141). Where the AI system is tested for its intended purpose outside a laboratory or otherwise simulated environment (Art. 3 (57)), providers must either comply with the AI Act's conditions for real-world testing such as compiling a real-world testing plan (Arts. 3 (53); 60) or conduct the testing in the context of a regulatory sandbox (Arts. 3 (55); 58-59), see \citet{buocz2023regulatory} for details.

\subsection{Research Exceptions for GPAI}
Due to the inconsistent usage of the terms AI system and AI model in the AI Act, the applicability of research exceptions for scientific (Art. 2(6)) and product-oriented research (Art. 2 (8)) to GPAI models is subject to further uncertainty. Both exceptions refer to AI models but do not specifically mention GPAI. Similar terminology is adopted in reference to some obligations associated with HRAI (Arts. 10; 15). In our view, the rest of the Act and the recitals suggest that GPAI is subject to distinct exceptions of its own.

The definition of a GPAI model includes its own product-oriented research exception. The exception applies to (GP)AI models that are used for research, development, or prototyping activities before they are placed on the market (Art. 3 (63)). The recitals affirm that compliance obligations for the providers of GPAI models do not extend to the persons who develop and use them for scientific purposes (Rec. 109).

The AI Act's recitals also appear to carve out a distinct research exception for GPAI models with systemic risks.\footnote{\textbf{systemic risk} means a risk that is specific to the high-impact capabilities of general-purpose AI models, having
a significant impact on the Union market due to their reach, or due to actual or reasonably foreseeable negative effects
on public health, safety, public security, fundamental rights, or the society as a whole, that can be propagated at scale
across the value chain (Art 3 (65))} The obligations for such systems should not cover GPAI models used before placing them on the market for the sole purpose of research, development, and prototyping activities (Rec. 97). It is notable that the research exception for GPAI with systemic risk is present only in the recitals, which, unlike the text of the Act, are non-binding \cite{klimas2008law}. 

Similarly to the research exception discussed above, the research exceptions for GPAI models are vague. What does ``research before placing on the market'' indicate? When is a GPAI model used for research? Again, uploading a model to a standard ML repository like GitHub or a service like Hugging Face/ Colab can be seen as part of research. When we refer to the narrow research definition from the Oxford Dictionary mentioned above, one can also question this assumption (see also section \ref{sec:SR}). Therefore, the scope of the exception is, again, unclear. Furthermore, in light of the narrow definition of sole scientific research for HRAI (Art. 2(6), Rec. 25), how can prototyping activity represent the sole purpose-use of a GPAI model with a systemic risk?

\subsection{Exceptions for Open Source AI}
The AI Act acknowledges the role of free and open source licenses for software and data for research and innovation (Rec. 102). The Act does not apply to AI systems released under free and open source licenses (Art. 2(12)). Could this exception protect researchers who want to publish their AI models on GitHub? It should be noted that not all models published on the platform are released under a free and open source license. However, more diligent habits of relying on free and open source licensing will not mitigate AI researchers' risk of becoming subject to the AI Act's obligations: the exception for free and open source AI does not apply to HRAI systems (Art. 2 (12)) or GPAI models with systemic risks (Art. 54 (6)). Moreover, making a GPAI model available under an open source leads to only partial exemption from GPAI provider obligations (Art. 53 (2)). Additionally, correct licensing on GitHub is currently rather challenging for the computer science community \citep{vendome2017license,wolter2023open}.

\subsection{No Exceptions for Prohibited AI?}
The legal consequences for prohibited uses of AI are triggered by placing the systems on the market, putting them into service, or \textit{the use} of the said systems (Art. 5). In other words, also the deployment of a such an AI system placed on the market by another provider is prohibited (Art. 3 (4)). Does the exception of Art 2(6) permit the development and putting into service of prohibited AI systems for the purposes of scientific research? After all, research on existing prohibited systems may be necessary for the development of more trustworthy AI systems. It should be noted, that upon redlining the prohibited AI system, the EU legislator prioritized the high level of protection for health, safety and fundamental rights against the Acts' parallel objectives of promoting the uptake of human-centric and trustworthy AI or support to innovation (Rec. 1). This speaks in favour of a narrow interpretation of the exception for scientific research with respect to prohibited AI systems, underscoring the need to ascertain that the systems are specifically developed for the sole scientific purpose (Art. 3 (6)) In-house research prohibited systems developed elsewhere could trigger deployers' liability under the AI Act. Arguably, the AI Act leaves a very narrow leeway to conduct research on prohibited AI systems, such as their examination.

\subsection{Summary of Exceptions: Substantial Legal Uncertainty}
As we have shown, at first glance, the AI Act features several exceptions in favour of AI research. However, taking a closer look, we showed that the exceptions might not be as broad as expected, due to their disconnect from the research practices of the field of Machine Learning. Therefore, we argue that the exceptions leave researchers with substantial legal uncertainty and risk suppressing the established practices of publishing AI research artifacts.

\section{Alternative Views}
Our viewpoint that researchers are covered by the AI Act, can be challenged by four different arguments.\\
Firstly, freedom of science is a fundamental right of the European Union (Art. 13 \cite{CFREU}). The AI Act should not affect scientific research beyond what is specified in the Act (Rec. 25). The fact that the AI Act does not refer to the publication of AI systems and models in the course of academic research (Art. 2(6)) does not mean that it applies to this scientific research practice. %
We do not agree with this viewpoint since, as explained above, the AI Act directly includes research exceptions, and thus protects the freedom of science. Our position is that these exceptions do not account for key practices in AI research. \\
The second option would be to argue that publishing AI models or systems by scientific researchers does not qualify as making them available to the public because this does not represent a ``commercial activity'' (Art. 3 (10)). The Blue Guide clarifies the interpretation of EU's product harmonization legislation, which the AI Act is part of (Rec. 46)\cite{Bluebook}. Commercial activity refers to providing goods in a business-related context, the presence of which is evaluated on a case-by-case basis, taking into account the regularity of supplies, the characteristics of the supplies, the characteristics of the producer, the intentions of the supplies, etc. Also, non-profit organizations may be considered as carrying out commercial activities if they operate in such a context. However, occasional supplies by charities or hobbyists are not deemed to be part of the business context \citep{Commission_2022}. The assessment of the presence of 'commercial activity' is difficult, especially in collaborations with heterogeneous stakeholders, \cite{finck2026eu}. In our view, the objectives of the AI Act call for a more expansive reading of the term "commercial". Whereas EU product legislation typically concerns itself with risks to health and safety, the AI Act also covers risks to fundamental rights. Consistently with this objective, the AI Act established obligations not only to providers of AI but also to deployers of AI systems. AI providers and deployers encompass  natural or legal persons, public authorities, agencies, or other bodies (Art. 3(3)\&(4)). This suggests that a researcher could also be a provider \citep{smuha2021eu}. Moreover, many of the high-risk AI systems fall into the area of public governance. A narrow interpretation of the term "commercial"  that would exclude public actors, including research institutions, from the scope of the Act on the basis of them operating non-commercially would vacate the purpose of the AI Act. On this basis, the AI Act's concept of "commercial activity" is very broad and encompasses also professional activity. This would be consistent with the fact that the AI Act offers an exception for personal non-professional activity only to AI deployers, and not providers (Art. 3(4)). 

Third, one can also argue that the EU AI Act can be seen, at least partly, as product safety law \citep{almada2023eu}. Product safety law applies to finished products \cite{Bluebook}. Products are outputs of industries and not of research. Thus, while the intention of the EU AI Act is to regulate products, science would be excluded from the scope. We agree that the AI Act focuses on products. Many of its rules on high-risk systems and GPAI models focus on industry-related output. Nevertheless, it is important to note that the AI Act addresses technology's impact on individuals, such as their fundamental rights \citep{kusche2024possible, almada2023eu, buocz2023regulatory, wernick2024impact}. Moreover, it may be impossible to define when an AI system represents a finished product, since, by definition, ``may exhibit adaptiveness after deployment" (Art 3(1)). These aspects speak in favour of seeing the AI Act to extend beyond the scope of traditional product safety regulation, and therefore cover also the use and dissemination of research outputs.
Fourth, one could also argue in favour regulating AI research. We fully agree that scientific freedom should not prevail where it poses threats to health, safety or fundamental rights, for example. There can be scenarios where it is undesirable to provide uncontrolled access to or publish certain AI systems, even in the context of scientific research. However, the rules of the AI act are neither tailored for filtering our such AI systems in the process of research, nor establishing binding rules for ethical AI research. Rather, its non-binding recitals meekly state that any R\&D activity should be carried out in accordance with recognised ethical and professional standards for scientific research (rec 25). If the EU legislator aims to control the dissemination of ``forbidden knowledge'' \citep{hagendorff2021forbidden}, the Act's research exceptions should be accompanied by safeguards fitting the process of AI research instead of exposing scientists to obligations tailored for product manufacturers. Unclear rules may lead to situations where risk-averse researchers refrain from publishing also where it does not breach the AI Act.

\section{Call for Action: Legal Certainty is Needed for Science}
\label{sec:Certanty}

What at first glance seems like a range of exceptions in favour of research is a collection of norms with complex systematics, such as differences between research exceptions in favour of high-risk AI and GPAI. The rules also contain numerous exceptions-of-exceptions and terminological ambiguities. As a result, we argue that the AI Act does not align well with AI research practices and exposes AI researchers to compliance obligations --- and liabilities. The legal uncertainty may hinder the publishing of AI artifacts, especially on behalf of AI researchers working at commercial companies. 

Three interpretations could bring the AI Act closer in alignment with AI research practices, albeit offering limited legal certainty. First, the AI Act poses compliance obligations to the providers of high-risk AI \textit{systems}. Consequently, the Act can be interpreted as not applying to the making available of high-risk AI \textit{models} (Art. 2(6)). AI researchers could attempt to publish only AI models (for example, the weights of a neural network), as opposed to AI systems (see above Section \ref{sec:ModelSystem} in the appendix for the difference).\\
The second option is to stress that the model or a system was made available for the intended purpose of scientific research. Under the AI Act, the concept of \textit{intended purpose} refers to the use for which an AI \textit{system} is intended by the provider, including the specific context and conditions of use as specified in the information supplied by the provider (Art. 3(12)). The intended purpose is determined by the provider under conditions, which can be reasonably foreseen \citep{Commission_2022}. The AI system's intended purpose determines whether the system is deemed high-risk (Art. 6, Annex III). The intended purpose also qualifies the nature and scope of the provider's compliance measures (Section 2; \citep{Commission_2022}). Consequently, the provider is not liable for the consequences of a third party making significant changes either to a high-risk system (Art. 25 (1)(b)) or the \textit{intended purpose} of a previously non-high-risk system (Art. 25 (1)(c)). To protect consumers' interests, it could make sense to label these ``as pre-compliant research publications." It should be also noted that the AI Act requires providers to account for reasonably foreseeable risks and misuses. The interpretation does not hold where an AI system or a model poses obvious risks to fundamental rights, health, or safety. Moreover, this interpretation may not suffice for GPAI models. In addition, with such a ``purpose-limitation," a GPAI model could not qualify as an open source GPAI under the AI Act\cite{Commission_2025}.\\
Third, the AI Act may permit license-based limitations to the use of AI systems and models. The recent Commission Guidelines point to such a possibility: the licensors of open-source GPAI models may establish more specific, safety-oriented terms that reasonably restrict usage in applications or domains where such use would pose a significant risk to public safety, security, or fundamental rights\cite{Commission_2025}. However, the strength and scope of disclaimers, contracts, and licenses in limiting researchers liability is uncertain. 

How likely is the AI Act influencing research?  We believe the GDPR offers a historical precedent from which we would estimate a high baseline probability that the AI Act will affect researchers. In general, the AI Act's implementation will unfold gradually before its full consequences become apparent to practitioners. The GDPR (which has a different legal regime in the background) provides an instructive parallel: its advent posed significant difficulties for research, especially in biomedical contexts \cite{Peloquin2020GDPR}. Crucially, this influence did not materialise overnight. Studies eventually highlighted challenges for cancer research \cite{Lawlor2023GDPR} and data science \cite{Greene2019GDPR}. Additionally, in clinical settings, work \cite{Davey2022GDPR} shows that only around one-sixth of doctors reported being ``very familiar'' with GDPR provisions even well after its enactment. This illustrates that awareness of a law's research implications builds slowly. In parallel to GDPR, we estimate that the probability that the AI Act influences research could be higher than 50\%. Compared to the GDPR, the AI Act does not impose all of its obligations at once but follows a staged applicability. This makes it even more difficult for researchers to become familiar with the obligations in time.

Echoing other authors \cite{finck2025search, latil2025regulating}, we hold the position that the EU should either revise the AI Act or issue an authoritative interpretation of the Act to provide legal certainty for AI research. 
Recently, EU policy has displayed signs of backtracking from heavy regulation of technology to spearhead competitiveness and further investments in the R\&D funding for AI \cite{Draghi}\cite{Virkkunen}. In late 2025, European Commission issued a ``Digital Omnibus" proposal to simplify the implementation of the Act. Whereas proposal recognized the need to issue guidelines on practical application the research exceptions of the AI act (Art. 2(6) and 2 (8)), the urgency was motivated by research in medical AI\cite{OmnibusAI}.

Within the EU, we see a need for ML researchers to engage more closely with policymakers. Our paper highlights how overlooking the needs and practices of ML researchers in AI policymaking can lead to undesirable outcomes. We see path dependency as a reason: EU legislators are faced with drafting research exceptions when regulating various technologies. In the case of AI, EU legislators did not sufficiently acknowledge that the object of regulation (AI systems, GPAI models) and scientific output may overlap. We argue that initiatives at the European level, like the Codes of Practice drafting \cite{EC_GPAI_CoP, EC_AIGenContent_CoP}, the EU AI Act scientific panel \cite{EC_ScientificPanel}, and the advisory forum \cite{EC_AdvisoryForum}, offer compelling possibilities for researchers to impact policymaking inside the EU.
Outside the EU, it is highly likely that, through the Brussels effect (see above), the AI Act will influence policymaking around the globe. Therefore, our paper suggests that ML researchers should engage in the policymaking process as soon as possible to avoid challenges later in time.

In our view, the AI researcher community represents a distinct stakeholder group, whose needs should not be overlooked by the EU. Its legislative drafting, revisions, and interpretations should account for the actual practices of AI research.
In the absence of case-law, legal uncertainty prevails. 
Gaining clarity should not depend on an unfortunate researcher being held liable for breaching the AI Act, and dragging them through the European judicial system. Albeit research organizations are an unlikely enforcement priority, the uncertainty has undesirable effects. Without clear rules for researchers, principal investigators, and more importantly, legal departments overseeing high stakes research projects will err on the side of caution with respect to publishing AI artifacts to the detriment of science. \\
First, the AI Act should be revised, or an authoritative interpretation should be issued to confirm scope of the AI researchers' freedom to publish. The European Commission should affirm a) how AI models differ from AI systems, b) under which conditions academic publishing of artifacts triggers the AI Act's obligations, c) the validity of disclaimers, contractual conditions and licensing conditions limiting the subsequent use of published AI artifacts to scientific research, and d) the scope of research exceptions with respect to prohibited AI, HRAI and GPAI. Second, where researchers and research institutions are obliged to comply with the AI Act, they should enjoy the same privileges under the Act as do SME's and start-ups do (
Art. 11 (1); 58 (2)-(3), 62).
Finally, the R\&D exceptions and possible the accommodations for academic publishing of AI should not enable commercial actors to skirt from provider's obligations \citep{colonna2023ai}.

\vspace{-0.2cm}
\section{Conclusion}
AI researchers taking part in academic conferences are required to publish their AI systems or models. The act of publishing qualifies as making them available on the European market and can trigger the applicability of the AI Act. 
Generally, AI research may concern AI categories protected by the AI Act. The AI Act's complex research exceptions do not provide a safe haven for publishing such AI systems or models. The AI Act could be interpreted to not apply to high-risk AI models and to permit AI researchers to limit their liability with disclaimers. However, the AI research community would need more legal certainty. In the current state, the Act risks chilling academic research practices at the expense of scientific research and innovation, potentially hindering the development more trustworthy AI, thus undermining values supported by the AI Act.

\section*{Authorship contributions}
Alina Wernick led the conceptualization of the research and doctrinal analysis, including Figure 1. Kristof Meding contributed to the doctrinal analysis, and was responsible for bridging the interface between law and machine learning, including Figure 2.

\section*{Acknowledgements}
We would like to thank Tharushi Abeynayaka, Rabanus Derr, Henrik Nolte, Sanidhya Rao, Mara Seyfert, Alan Medlar, Vittorio Franzese and Irene Rocchesso for their detailed comments on our manuscript. Finally, we would also like to thank our anonymous reviewer, whose suggestions significantly improved the manuscript.

Alina Wernick and Kristof Meding were supported by the Carl Zeiss Foundation through the CZS Center for AI and Law. Additionally, both are members of the Machine Learning Cluster of Excellence, funded by the Deutsche Forschungsgemeinschaft (DFG, German Research Foundation) under Germany’s Excellence Strategy – EXC number 2064/1 – Project number 390727645.

\bibliography{bibliography.bib}

\begin{thebibliography}{74}
\providecommand{\natexlab}[1]{#1}
\providecommand{\url}[1]{\texttt{#1}}
\expandafter\ifx\csname urlstyle\endcsname\relax
  \providecommand{\doi}[1]{doi: #1}\else
  \providecommand{\doi}{doi: \begingroup \urlstyle{rm}\Url}\fi

\bibitem[Almada \& Petit(2025)Almada and Petit]{almada2023eu}
Almada, M. and Petit, N.
\newblock The eu ai act: Between the rock of product safety and the hard place of fundamental rights.
\newblock \emph{Common Market Law Review}, 62\penalty0 (1), 2025.

\bibitem[Almazrouei et~al.(2023)Almazrouei, Alobeidli, Alshamsi, Cappelli, Cojocaru, Debbah, Goffinet, Hesslow, Launay, Malartic, et~al.]{almazrouei2023falcon}
Almazrouei, E., Alobeidli, H., Alshamsi, A., Cappelli, A., Cojocaru, R., Debbah, M., Goffinet, {\'E}., Hesslow, D., Launay, J., Malartic, Q., et~al.
\newblock The falcon series of open language models.
\newblock \emph{arXiv preprint arXiv:2311.16867}, 2023.

\bibitem[Artificial Intelligence~Act(2024)]{AIA}
Artificial Intelligence~Act, R. E.~., 2024.
\newblock URL \url{http://data.europa.eu/eli/reg/2024/1689/oj}.

\bibitem[Biega \& Finck(2021)Biega and Finck]{biega2021reviving}
Biega, A.~J. and Finck, M.
\newblock Reviving purpose limitation and data minimisation in data-driven systems.
\newblock \emph{arXiv preprint arXiv:2101.06203}, 2021.

\bibitem[Bogucki et~al.(2022)Bogucki, Engler, Perarnaud, and Renda]{bogucki2022ai}
Bogucki, A., Engler, A., Perarnaud, C., and Renda, A.
\newblock The ai act and emerging eu digital acquis.
\newblock \emph{Overlaps, gaps and}, 2022.

\bibitem[Bosoer et~al.(2023)Bosoer, Cantero~Gamito, and Rubio-Marin]{bosoer2023non}
Bosoer, L., Cantero~Gamito, M., and Rubio-Marin, R.
\newblock Non-discrimination and the ai act.
\newblock \emph{Law and Digitalization. Arazandi}, 2023.

\bibitem[Bradford(2020)]{bradford2020brussels}
Bradford, A.
\newblock \emph{The Brussels effect: How the European Union rules the world}.
\newblock Oxford University Press, 2020.

\bibitem[Brown et~al.(2020)Brown, Mann, Ryder, Subbiah, Kaplan, Dhariwal, Neelakantan, Shyam, Sastry, Askell, et~al.]{brown2020language}
Brown, T., Mann, B., Ryder, N., Subbiah, M., Kaplan, J.~D., Dhariwal, P., Neelakantan, A., Shyam, P., Sastry, G., Askell, A., et~al.
\newblock Language models are few-shot learners.
\newblock \emph{Advances in neural information processing systems}, 33:\penalty0 1877--1901, 2020.

\bibitem[Bublitz et~al.(2024)Bublitz, Moln{\'a}r-G{\'a}bor, and Soekadar]{bublitz2024implications}
Bublitz, C., Moln{\'a}r-G{\'a}bor, F., and Soekadar, S.~R.
\newblock Implications of the novel eu ai act for neurotechnologies.
\newblock \emph{Neuron}, 112\penalty0 (18):\penalty0 3013--3016, 2024.

\bibitem[Buocz et~al.(2023)Buocz, Pfotenhauer, and Eisenberger]{buocz2023regulatory}
Buocz, T., Pfotenhauer, S., and Eisenberger, I.
\newblock Regulatory sandboxes in the ai act: reconciling innovation and safety?
\newblock \emph{Law, Innovation and Technology}, 15\penalty0 (2):\penalty0 357--389, 2023.

\bibitem[CFREU(2000)]{CFREU}
CFREU.
\newblock Charter of fundamental rights of european union, 2000.
\newblock URL \url{https://www.europarl.europa.eu/charter/pdf/text_en.pdf}.

\bibitem[Chiappetta(2023)]{chiappetta2023navigating}
Chiappetta, A.
\newblock Navigating the ai frontier: European parliamentary insights on bias and regulation, preceding the ai act.
\newblock \emph{Internet Policy Review}, 12\penalty0 (4), 2023.

\bibitem[Chynoweth(2008)]{chynoweth2008legal}
Chynoweth, P.
\newblock Legal research.
\newblock \emph{Advanced research methods in the built environment}, 1, 2008.

\bibitem[Colonna(2023)]{colonna2023ai}
Colonna, L.
\newblock The ai act’s research exemption: A mechanism for regulatory arbitrage?
\newblock In \emph{YSEC Yearbook of Socio-Economic Constitutions 2023: Law and the Governance of Artificial Intelligence}, pp.\  51--93. Springer, 2023.

\bibitem[Commission(2021)]{IAAI2021}
Commission, E.
\newblock Commission staff working document impact assessment accompanying the proposal for a regulation of the european parliament and of the council laying down harmonised rules on artificial intelligence (artificial intelligence aact) and amending certain union legislative acts, 2021.
\newblock URL \url{https://ec.europa.eu/newsroom/dae/redirection/document/75792}.

\bibitem[Commission(2022)]{Bluebook}
Commission, E.
\newblock The ‘blue guide’ on the implementation of eu product rules, 2022.
\newblock URL \url{https://eur-lex.europa.eu/legal-content/EN/TXT/PDF/?uri=OJ:C:2022:247:FULL}.

\bibitem[Commission(2022/C)]{Commission_2022}
Commission, E.
\newblock \emph{The `Blue Guide´ on the implementation of EU product rules 2022}, volume~65.
\newblock The European Union, 2022/C.

\bibitem[Commission(2025{\natexlab{a}})]{Commission_2025}
Commission, E.
\newblock \emph{Approval of the content of the draft Communication from the Commission – Guidelines on the scope of the obligations for general-purpose AI models established by Regulation (EU) 2024/1689 (AI Act)}, volume C(2025) 5045 final.
\newblock The European Union, 2025{\natexlab{a}}.

\bibitem[Commission(2025{\natexlab{b}})]{EUAIDigitalStrategy2025}
Commission, E.
\newblock European research development and deployment of ai, 2025{\natexlab{b}}.
\newblock URL \url{https://digital-strategy.ec.europa.eu/en/policies/european-ai-research}.

\bibitem[Commission(2025{\natexlab{c}})]{OmnibusAI}
Commission, E.
\newblock Proposal for a regulation of the european parliament and of the council amending regulations (eu) 2024/1689 and (eu) 2018/1139 as regards the simplification of the implementation of harmonised rules on artificial intelligence (digital omnibus on ai), 2025{\natexlab{c}}.
\newblock URL \url{https://ec.europa.eu/newsroom/dae/redirection/document/121744}.

\bibitem[Davey et~al.(2022)Davey, O'Donnell, Maher, McMenamin, McAnena, Kerin, Miller, and Lowery]{Davey2022GDPR}
Davey, M.~G., O'Donnell, J. P.~M., Maher, E., McMenamin, C., McAnena, P.~F., Kerin, M.~J., Miller, N., and Lowery, A.~J.
\newblock General data protection regulations (2018) and clinical research: perspectives of patients and doctors in an {I}rish university teaching hospital.
\newblock \emph{Irish Journal of Medical Science (1971 -)}, 191\penalty0 (4):\penalty0 1513--1519, 2022.
\newblock \doi{10.1007/s11845-021-02789-8}.

\bibitem[De~Vries(2021)]{de2021researcher}
De~Vries, K.
\newblock A researcher’s guide for using personal data and non-personal data surrogates: Synthetic data and data of deceased people.
\newblock \emph{De Lege}, 2021.

\bibitem[Deck et~al.(2024)Deck, Schoem{\"a}cker, Speith, Sch{\"o}ffer, K{\"a}stner, and K{\"u}hl]{deck2024mapping}
Deck, L., Schoem{\"a}cker, A., Speith, T., Sch{\"o}ffer, J., K{\"a}stner, L., and K{\"u}hl, N.
\newblock Mapping the potential of explainable artificial intelligence (xai) for fairness along the ai lifecycle.
\newblock \emph{arXiv preprint arXiv:2404.18736}, 2024.

\bibitem[Draghi(2024)]{Draghi}
Draghi, M.
\newblock The future of european competitiveness. part a.
\newblock Technical report, European Commission, 2024.
\newblock URL \url{https://commission.europa.eu/document/download/97e481fd-2dc3-412d-be4c-f152a8232961_en?filename=The%20future%20of%20European%20competitiveness%20_%20A%20competitiveness%20strategy%20for%20Europe.pdf}.

\bibitem[Dubey et~al.(2024)Dubey, Jauhri, Pandey, Kadian, Al-Dahle, Letman, Mathur, Schelten, Yang, Fan, et~al.]{dubey2024llama}
Dubey, A., Jauhri, A., Pandey, A., Kadian, A., Al-Dahle, A., Letman, A., Mathur, A., Schelten, A., Yang, A., Fan, A., et~al.
\newblock The llama 3 herd of models.
\newblock \emph{arXiv preprint arXiv:2407.21783}, 2024.

\bibitem[{European Commission}({\natexlab{a}})]{EC_AIGenContent_CoP}
{European Commission}.
\newblock Code of practice on {AI}-generated content.
\newblock \url{https://digital-strategy.ec.europa.eu/en/policies/code-practice-ai-generated-content}, {\natexlab{a}}.
\newblock Accessed: 12 May 2026.

\bibitem[{European Commission}({\natexlab{b}})]{EC_AdvisoryForum}
{European Commission}.
\newblock European commission launches call for applications to join the {AI} act advisory forum.
\newblock \url{https://digital-strategy.ec.europa.eu/en/funding/european-commission-launches-call-applications-join-ai-act-advisory-forum}, {\natexlab{b}}.
\newblock Accessed: 12 May 2026.

\bibitem[{European Commission}({\natexlab{c}})]{EC_GPAI_CoP}
{European Commission}.
\newblock General-purpose {AI} code of practice.
\newblock \url{https://digital-strategy.ec.europa.eu/en/policies/contents-code-gpai}, {\natexlab{c}}.
\newblock Accessed: 12 May 2026.

\bibitem[{European Commission}({\natexlab{d}})]{EC_ScientificPanel}
{European Commission}.
\newblock Commission seeks experts for the {AI} scientific panel.
\newblock \url{https://digital-strategy.ec.europa.eu/en/news/commission-seeks-experts-ai-scientific-panel}, {\natexlab{d}}.
\newblock Accessed: 12 May 2026.

\bibitem[Finck(2025)]{finck2025search}
Finck, M.
\newblock In search of the lost research exemption: Reflections on the ai act, 2025.

\bibitem[Finck(2026)]{finck2026eu}
Finck, M.
\newblock \emph{The EU Artificial Intelligence Act: A Commentary}.
\newblock Oxford University Press, 2026.

\bibitem[Gils(2024)]{gils_detailed_2024}
Gils, T.
\newblock A {Detailed} {Analysis} of {Article} 50 of the {EU}'s {Artificial} {Intelligence} {Act}, 2024.

\bibitem[Gorwa \& Veale(2024)Gorwa and Veale]{gorwa2024moderating}
Gorwa, R. and Veale, M.
\newblock Moderating model marketplaces: Platform governance puzzles for ai intermediaries.
\newblock \emph{Law, Innovation and Technology}, 16\penalty0 (2):\penalty0 341--391, 2024.

\bibitem[Greene et~al.(2019)Greene, Shmueli, Ray, and Fell]{Greene2019GDPR}
Greene, T., Shmueli, G., Ray, S., and Fell, J.
\newblock Adjusting to the {GDPR}: The impact on data scientists and behavioral researchers.
\newblock \emph{Big Data}, 7\penalty0 (3):\penalty0 140--162, 2019.
\newblock \doi{10.1089/big.2018.0176}.

\bibitem[Guo et~al.(2025)Guo, Yang, Zhang, Song, Zhang, Xu, Zhu, Ma, Wang, Bi, et~al.]{guo2025deepseek}
Guo, D., Yang, D., Zhang, H., Song, J., Zhang, R., Xu, R., Zhu, Q., Ma, S., Wang, P., Bi, X., et~al.
\newblock Deepseek-r1: Incentivizing reasoning capability in llms via reinforcement learning.
\newblock \emph{arXiv preprint arXiv:2501.12948}, 2025.

\bibitem[Hacker et~al.(2023)Hacker, Engel, and Mauer]{hacker2023regulating}
Hacker, P., Engel, A., and Mauer, M.
\newblock Regulating chatgpt and other large generative ai models.
\newblock In \emph{Proceedings of the 2023 ACM Conference on Fairness, Accountability, and Transparency}, pp.\  1112--1123, 2023.

\bibitem[Hagendorff(2021)]{hagendorff2021forbidden}
Hagendorff, T.
\newblock Forbidden knowledge in machine learning reflections on the limits of research and publication.
\newblock \emph{AI \& Society}, 36\penalty0 (3):\penalty0 767--781, 2021.

\bibitem[Helberger et~al.(2024)Helberger, Naudts, and Piasecki]{Amsterdampaper}
Helberger, N., Naudts, L., and Piasecki, S.
\newblock The amsterdam paper: Recommendations for the technical finalisation of the regulation of gpai in the ai act, 2024.

\bibitem[Huang et~al.(2008)Huang, Mattar, Berg, and Learned-Miller]{huang2008labeled}
Huang, G.~B., Mattar, M., Berg, T., and Learned-Miller, E.
\newblock Labeled faces in the wild: A database forstudying face recognition in unconstrained environments.
\newblock In \emph{Workshop on faces in'Real-Life'Images: detection, alignment, and recognition}, 2008.

\bibitem[IAPP(2024)]{IAPP}
IAPP.
\newblock Global ai law and policy tracker, 2024.
\newblock URL \url{https://iapp.org/media/pdf/resource_center/global_ai_law_policy_tracker.pdf}.

\bibitem[Kazim et~al.(2023)Kazim, G{\"u}{\c{c}}l{\"u}t{\"u}rk, Almeida, Kerrigan, Lomas, Koshiyama, Hilliard, and Trengove]{kazim2023proposed}
Kazim, E., G{\"u}{\c{c}}l{\"u}t{\"u}rk, O., Almeida, D., Kerrigan, C., Lomas, E., Koshiyama, A., Hilliard, A., and Trengove, M.
\newblock Proposed eu ai act—presidency compromise text: select overview and comment on the changes to the proposed regulation.
\newblock \emph{AI and Ethics}, 3\penalty0 (2):\penalty0 381--387, 2023.

\bibitem[Kilian et~al.(2025)Kilian, J{\"a}ck, and Ebel]{kilian2025european}
Kilian, R., J{\"a}ck, L., and Ebel, D.
\newblock European ai standards-technical standardization and implementation challenges under the eu ai act.
\newblock \emph{Available at SSRN 5155591}, 2025.

\bibitem[Klimas \& Vaiciukaite(2008)Klimas and Vaiciukaite]{klimas2008law}
Klimas, T. and Vaiciukaite, J.
\newblock The law of recitals in european community legislation.
\newblock \emph{ILSA J. Int'l \& Comp. L.}, 15:\penalty0 61, 2008.

\bibitem[Kosinski(2021)]{kosinski2021facial}
Kosinski, M.
\newblock Facial recognition technology can expose political orientation from naturalistic facial images.
\newblock \emph{Scientific reports}, 11\penalty0 (1):\penalty0 100, 2021.

\bibitem[Kusche(2024)]{kusche2024possible}
Kusche, I.
\newblock Possible harms of artificial intelligence and the eu ai act: fundamental rights and risk.
\newblock \emph{Journal of Risk Research}, pp.\  1--14, 2024.

\bibitem[Latil(2025)]{latil2025regulating}
Latil, A.
\newblock Regulating general-purpose ai models: A dual disruption.
\newblock \emph{THE ACADEMIC GUIDE TO AI ACT COMPLIANCE}, pp.\ ~29, 2025.

\bibitem[Lawlor(2023)]{Lawlor2023GDPR}
Lawlor, R.~T.
\newblock The impact of {GDPR} on data sharing for european cancer research.
\newblock \emph{The Lancet Oncology}, 24\penalty0 (1):\penalty0 6--8, 2023.
\newblock \doi{10.1016/S1470-2045(22)00653-2}.

\bibitem[Le~Scao et~al.(2023)Le~Scao, Fan, Akiki, Pavlick, Ili{\'c}, Hesslow, Castagn{\'e}, Luccioni, Yvon, Gall{\'e}, et~al.]{le2023bloom}
Le~Scao, T., Fan, A., Akiki, C., Pavlick, E., Ili{\'c}, S., Hesslow, D., Castagn{\'e}, R., Luccioni, A.~S., Yvon, F., Gall{\'e}, M., et~al.
\newblock Bloom: A 176b-parameter open-access multilingual language model, 2023.

\bibitem[Mann et~al.(2024)Mann, Cohen, and Minssen]{mann2024eu}
Mann, S.~P., Cohen, I.~G., and Minssen, T.
\newblock The eu ai act: Implications for us health care, 2024.

\bibitem[Mantelero(2024)]{mantelero2024fundamental}
Mantelero, A.
\newblock The fundamental rights impact assessment (fria) in the ai act: Roots, legal obligations and key elements for a model template.
\newblock \emph{Computer Law \& Security Review}, 54:\penalty0 106020, 2024.

\bibitem[Mantelero(2025, forthcoming)]{Mantelero2025}
Mantelero, A.
\newblock Pre-print: The human-centric approach in scientific research: the ai act and the new frontiers of research ethics.
\newblock In \emph{Data Privacy, Data Property, and Data Sharing: An Interdisciplinary Perspective for Port-pandemic Transitional Scientific Research}. Boca Raton, CRC Press, 2025, forthcoming.

\bibitem[Meding(2025)]{meding2025s}
Meding, K.
\newblock It's complicated. the relationship of algorithmic fairness and non-discrimination regulations in the eu ai act.
\newblock \emph{arXiv preprint arXiv:2501.12962}, 2025.

\bibitem[Meding \& Sorge(2024)Meding and Sorge]{meding2024constitutes}
Meding, K. and Sorge, C.
\newblock What constitutes a deep fake? the blurry line between legitimate processing and manipulation under the eu ai act.
\newblock \emph{arXiv preprint arXiv:2412.09961}, 2024.

\bibitem[Möllers(2020)]{legalmethods}
Möllers, T.
\newblock \emph{Legal Methods. How to work with legal arguments}.
\newblock C.H. Beck, 2020.

\bibitem[Nolte et~al.(2024)Nolte, Rateike, and Finck]{Nolte2024}
Nolte, H., Rateike, M., and Finck, M.
\newblock Robustness and cybersecurity in the eu artificial intelligence act.
\newblock In \emph{Generative AI and Law (GenLaw ’24) Workshop at 41 International Conference on Machine Learning, Vienna, Austria}, 2024.

\bibitem[Parliament(2024)]{Virkkunen}
Parliament, E.
\newblock Questionnaire to the commissioner-designate henna virkkunen executive vice-president for tech sovereignty, security and democracy, 2024.
\newblock URL \url{https://hearings.elections.europa.eu/documents/virkkunen/virkkunen_writtenquestionsandanswers_en.pdf}.

\bibitem[Parliament \& Council(2024)Parliament and Council]{AIAcomparison}
Parliament, E. and Council.
\newblock Proposal for a regulation laying down harmonized rules on artificial intelligence (artificial intelligence act) and amending certain union legislative acts, four columns, 2024.
\newblock URL \url{https://artificialintelligenceact.eu/wp-content/uploads/2024/01/AIA-Final-Draft-21-January-2024.pdf}.

\bibitem[Peloquin et~al.(2020)Peloquin, DiMaio, Bierer, and Barnes]{Peloquin2020GDPR}
Peloquin, D., DiMaio, M., Bierer, B., and Barnes, M.
\newblock Disruptive and avoidable: {GDPR} challenges to secondary research uses of data.
\newblock \emph{European Journal of Human Genetics}, 28\penalty0 (6):\penalty0 697--705, 2020.
\newblock \doi{10.1038/s41431-020-0596-x}.

\bibitem[Pineau et~al.(2021)Pineau, Vincent-Lamarre, Sinha, Larivi{\`e}re, Beygelzimer, d'Alch{\'e} Buc, Fox, and Larochelle]{pineau2021improving}
Pineau, J., Vincent-Lamarre, P., Sinha, K., Larivi{\`e}re, V., Beygelzimer, A., d'Alch{\'e} Buc, F., Fox, E., and Larochelle, H.
\newblock Improving reproducibility in machine learning research (a report from the neurips 2019 reproducibility program).
\newblock \emph{Journal of machine learning research}, 22\penalty0 (164):\penalty0 1--20, 2021.

\bibitem[Quezada-Tavarez et~al.(2022)Quezada-Tavarez, Dutkiewicz, and Krack]{quezada2022voicing}
Quezada-Tavarez, K., Dutkiewicz, L., and Krack, N.
\newblock Voicing challenges: Gdpr and ai research.
\newblock \emph{Open Research Europe}, 2\penalty0 (126):\penalty0 126, 2022.

\bibitem[Quintais(2024)]{quintais2024generative}
Quintais, J.~P.
\newblock Generative ai, copyright and the ai act.
\newblock \emph{Available at SSRN}, 2024.

\bibitem[Smuha et~al.(2021)Smuha, Ahmed-Rengers, Harkens, Li, MacLaren, Piselli, and Yeung]{smuha2021eu}
Smuha, N.~A., Ahmed-Rengers, E., Harkens, A., Li, W., MacLaren, J., Piselli, R., and Yeung, K.
\newblock How the eu can achieve legally trustworthy ai: a response to the european commission’s proposal for an artificial intelligence act.
\newblock \emph{Available at SSRN 3899991}, 2021.

\bibitem[Tambe et~al.(2019)Tambe, Cappelli, and Yakubovich]{tambe2019artificial}
Tambe, P., Cappelli, P., and Yakubovich, V.
\newblock Artificial intelligence in human resources management: Challenges and a path forward.
\newblock \emph{California Management Review}, 61\penalty0 (4):\penalty0 15--42, 2019.

\bibitem[Team et~al.(2023)Team, Anil, Borgeaud, Alayrac, Yu, Soricut, Schalkwyk, Dai, Hauth, Millican, et~al.]{team2023gemini}
Team, G., Anil, R., Borgeaud, S., Alayrac, J.-B., Yu, J., Soricut, R., Schalkwyk, J., Dai, A.~M., Hauth, A., Millican, K., et~al.
\newblock Gemini: a family of highly capable multimodal models.
\newblock \emph{arXiv preprint arXiv:2312.11805}, 2023.

\bibitem[Vendome et~al.(2017)Vendome, Bavota, Penta, Linares-V{\'a}squez, German, and Poshyvanyk]{vendome2017license}
Vendome, C., Bavota, G., Penta, M.~D., Linares-V{\'a}squez, M., German, D., and Poshyvanyk, D.
\newblock License usage and changes: a large-scale study on github.
\newblock \emph{Empirical Software Engineering}, 22:\penalty0 1537--1577, 2017.

\bibitem[Wachter(2024)]{wachter2024limitations}
Wachter, S.
\newblock Limitations and loopholes in the eu ai act and ai liability directives: what this means for the european union, the united states, and beyond.
\newblock \emph{Yale Journal of Law and Technology}, 26\penalty0 (3), 2024.

\bibitem[Wernick(2024)]{wernick2024impact}
Wernick, A.
\newblock Impact assessment as a legal design pattern—a “timeless way” of managing future risks?
\newblock \emph{Digital Society}, 3\penalty0 (2):\penalty0 29, 2024.

\bibitem[White \& Case(n.d.)White and Case]{whitecase}
White and Case.
\newblock Ai watch: Global regulatory tracker, n.d.
\newblock URL \url{https://www.whitecase.com/insight-our-thinking/ai-watch-global-regulatory-tracker}.

\bibitem[Wolter et~al.(2023)Wolter, Barcomb, Riehle, and Harutyunyan]{wolter2023open}
Wolter, T., Barcomb, A., Riehle, D., and Harutyunyan, N.
\newblock Open source license inconsistencies on github.
\newblock \emph{ACM Transactions on Software Engineering and Methodology}, 32\penalty0 (5):\penalty0 1--23, 2023.

\bibitem[Yan et~al.(2024)Yan, Sha, Zhao, Li, Martinez-Maldonado, Chen, Li, Jin, and Ga{\v{s}}evi{\'c}]{yan2024practical}
Yan, L., Sha, L., Zhao, L., Li, Y., Martinez-Maldonado, R., Chen, G., Li, X., Jin, Y., and Ga{\v{s}}evi{\'c}, D.
\newblock Practical and ethical challenges of large language models in education: A systematic scoping review.
\newblock \emph{British Journal of Educational Technology}, 55\penalty0 (1):\penalty0 90--112, 2024.

\bibitem[Yigit et~al.(2024)Yigit, Ferrag, Sarker, Maglaras, Chrysoulas, Moradpoor, and Janicke]{yigit2024critical}
Yigit, Y., Ferrag, M.~A., Sarker, I.~H., Maglaras, L.~A., Chrysoulas, C., Moradpoor, N., and Janicke, H.
\newblock Critical infrastructure protection: Generative ai, challenges, and opportunities.
\newblock \emph{arXiv preprint arXiv:2405.04874}, 2024.

\bibitem[Yordanova(2022)]{yordanova2022eu}
Yordanova, K.
\newblock The eu ai act-balancing human rights and innovation through regulatory sandboxes and standardization, 2022.

\bibitem[Zuiderveen~Borgesius et~al.(2024)Zuiderveen~Borgesius, Hacker, Baranowska, and Fabris]{zuiderveen2024non}
Zuiderveen~Borgesius, F., Hacker, P., Baranowska, N., and Fabris, A.
\newblock Non-discrimination law in europe, a primer. introducing european non-discrimination law to non-lawyers.
\newblock \emph{Introducing European non-discrimination law to non-lawyers (April 7, 2024)}, 2024.

\bibitem[Łabuz(2024)]{labuz_deep_2024}
Łabuz, M.
\newblock Deep fakes and the artificial intelligence act-an important signal or a missed opportunity?
\newblock \emph{Policy \& Internet}, July 2024.
\newblock Publisher: Wiley.

\end{thebibliography}
\bibliographystyle{icml2026}

\newpage
\appendix
\section{Appendix}
\subsection{Previous Literature}
\label{ref:previousLiterature}
Different aspects of AI regulations in context of the AI Act have already been studied, such as non-discrimination regulations \citep{deck2024mapping,meding2025s,
chiappetta2023navigating,bosoer2023non}, deepfake regulations \citep{gils_detailed_2024, meding2024constitutes, labuz_deep_2024}, 
the impact on fundamental rights \citep{mantelero2024fundamental, kusche2024possible}, research ethics \cite{Mantelero2025} or lobbying efforts \citep{wachter2024limitations}. 
Few studies analyse the AI Act's or its earlier drafts' research exceptions in detail \citep{smuha2021eu,colonna2023ai,kazim2023proposed, Amsterdampaper,bogucki2022ai, yordanova2022eu}. The Act's research exceptions are usually discussed in connection with another topic, such as copyright or data protection \citep{quintais2024generative,quezada2022voicing,de2021researcher} or their relevance for a specific application area or type of AI \citep{bublitz2024implications,mann2024eu,hacker2023regulating}. Even the most in-depth study on the AI Act's research exceptions \citep{colonna2023ai} gives limited attention to the problem of regular scientific research publication practices triggering the application of the AI Act. \

\subsection{Models vs. Systems in the AI Act}
\label{sec:ModelSystem}
The AI Act differentiates between AI systems and models. The AI Act's definition of an AI \textit{system} is very broad, covering a vast range of theoretical and applied AI research: ``machine-based system that is designed to operate with varying levels of autonomy and that may exhibit adaptiveness after deployment, and that, for explicit or implicit objectives, infers, from the input it receives, how to generate outputs such as predictions, content, recommendations, or decisions that can influence physical or virtual environments'' (Art. 3(1)). 

Besides AI systems, the AI act regulates GPAI \textit{models}, which are defined as ``AI model, including where such an AI model is trained with a large amount of data using self-supervision at scale, that displays significant generality and is capable of competently performing a wide range of distinct tasks regardless of the way the model is placed on the market and that can be integrated into a variety of downstream systems or applications'' (Art. 2(63)).

If the AI Act regulates high-risk AI \textit{systems}, can an AI \textit{model} developed by an AI researcher also fall under this definition? The challenge is that the AI Act defines AI system in a manner that does not correspond to the language adopted in the AI research community \citep{Nolte2024}.\footnote{To add to the confusion, some of the AI Act's obligations for high-risk systems refer to models that are part of those systems (Art. 15 \cite{Nolte2024}).} This exposes researchers to considerable legal uncertainty because publishing a system is more likely to trigger liability under the AI Act than publishing an AI model. Looking, for example, at the NeurIPS Call for Papers\footnote{\url{https://neurips.cc/Conferences/2025/CallForPapers}.} we note that an ML model is the learned algorithmic structure itself (e.g. weights of a neural network). A Machine Learning system also includes the hardware, libraries, etc. AI Models require the addition of further components, such as, for example, a user interface, to become AI systems (Recital 97; \cite{Nolte2024}). However, even if we assume a very narrow definition of an AI model, at least, research offering demos of their research output (for example, on a website) can be seen as an AI system.%
\newpage
\section{Figure AI Research Lifecycle}
\begin{strip}
    \centering
    \includegraphics[width=\linewidth]{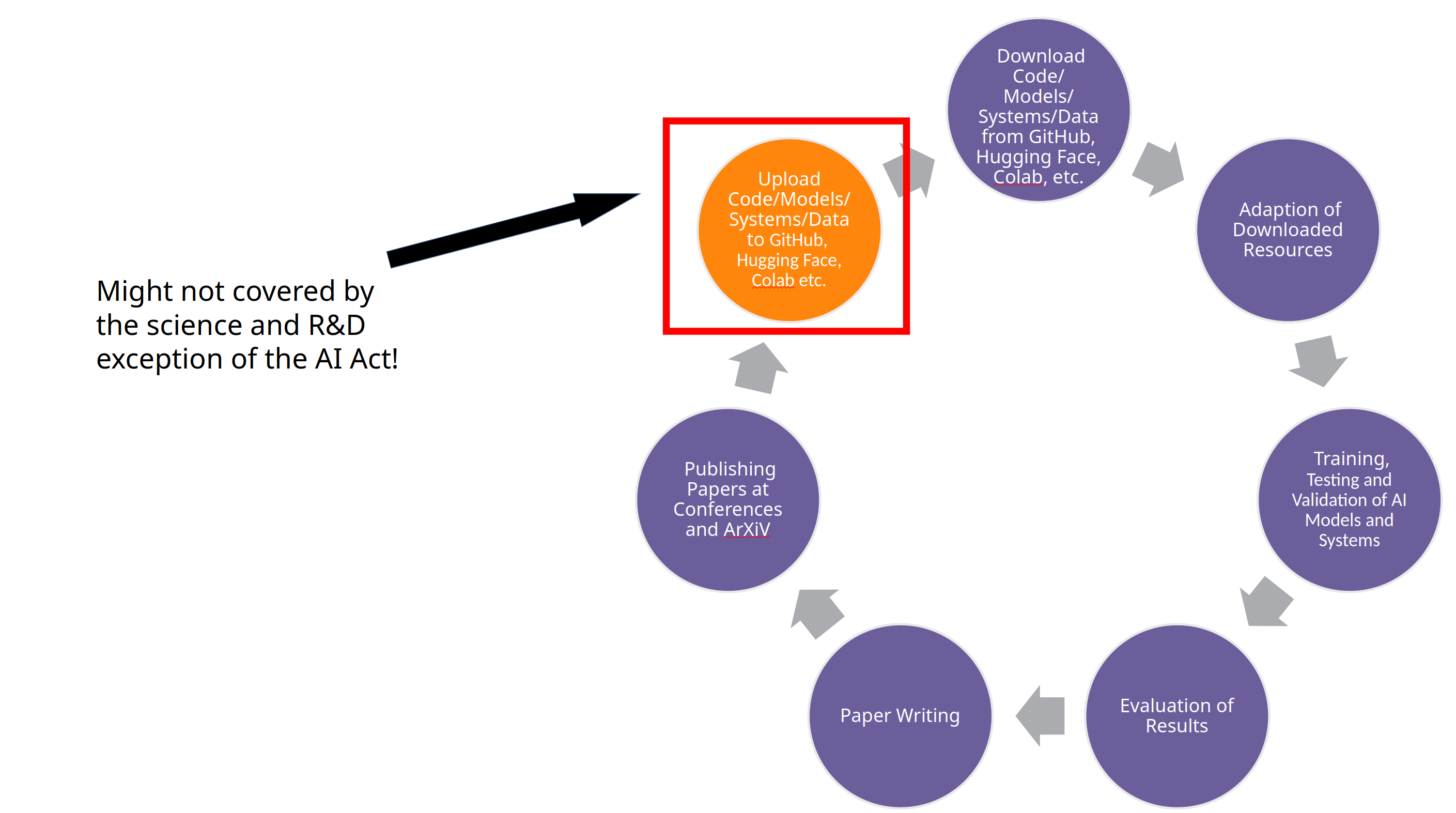}
    \captionof{figure}{AI research lifecycle: Researchers start by downloading code, systems, models, and data for their research. They adapt these resources to train, test, and validate their systems and models. Afterwards, the results are evaluated. Papers are written and published at conferences and on arXiv. Simultaneously, the code, models, and data are published to public repositories. We argue that publishing code, models, and data (orange) is not always covered by the science and R\&D exception of the AI Act, such that legal obligations of the AI Act might apply. In that case, uploading the model to certain repositories could turn the model into an AI system and thereby trigger the applicability of the AI Act. }
    \label{fig:ML_Workflow}
\end{strip}
\null

\end{document}